\documentclass[12pt]{article}

\usepackage[english]{babel}
\usepackage[utf8x]{inputenc}
\usepackage[T1]{fontenc}
\usepackage{physics}
\usepackage{float}
\usepackage{parskip}
\usepackage[affil-it]{authblk}
\usepackage{geometry}

\usepackage{amsmath}
\usepackage{graphicx}
\usepackage[colorinlistoftodos]{todonotes}
\usepackage[colorlinks=true, allcolors=blue]{hyperref}
\usepackage{xfrac}
\usepackage{units}
\usepackage{amssymb}
\usepackage{dsfont}
\usepackage{epstopdf}
\usepackage{subfig}
\usepackage{tikz}
\usetikzlibrary{calc}
\usepackage{nicefrac}
\usepackage{mathtools}
\usepackage{bm}
\usepackage{epstopdf}

\DeclarePairedDelimiter{\floor}{\lfloor}{\rfloor}

\title{Coherent states of the two-dimensional non-separable supersymmetric Morse potential}
\author[1,2]{James Moran \thanks{james.moran@umontreal.ca}}
\author[2,3]{V\'eronique Hussin \thanks{hussin@dms.umontreal.ca}}
\affil[1]{D\'epartement de physique, Universit\'e de Montr\'eal, Montr\'eal, Qu\'ebec, H3C 3J7, Canada}
\affil[2]{Centre de recherches math\'ematiques, Universit\'e de Montr\'eal, Montr\'eal, Qu\'ebec, H3C 3J7, Canada}
\affil[3]{D\'epartement de math\'ematiques et de statistique, Universit\'e de Montr\'eal, Montr\'eal,
Qu\'ebec, H3C 3J7, Canada}
\date{}

\begin{document}

\maketitle
\begin{abstract}
Supersymmetry is a technique that allows us to extract information about the states and spectra of quantum mechanical systems which may otherwise be unsolvable. In this paper we reconstruct Ioffe's set of states for the singular non-separable two-dimensional Morse potential using supersymmetry from a non-degenerate set of states constructed for the initial separable Morse Hamiltonian. We define generalised coherent states, compute their uncertainty relations, and we find that the singularity in the partner Hamiltonian significantly affects the localisation of the coherent state wavefunction.
\end{abstract}
\newpage
\tableofcontents
%%%%%%%%%%%%%%%%%%%%%%%%%%%%%%%%%%%%%%%%%%%%%%%%%%%%%%%%%%%%%%%%%%%%%%%%%%%%%%%%%%%%%%%

%%%%%%%%%%%%%%%%%%%%%%%%%%%%%%%%%%%%%%%%%%%%%%%%%%%%%%%%%%%%%%%%%%%%%%%%%%%%%%%%%%%%%%%

\section{Introduction}
Supersymmetric quantum mechanics is a powerful tool in solving new quantum problems by their mathematical relationship to known solved problems \cite{WITTEN1981513, 10.4310/jdg/1214437492}. When two systems are related through supersymmetry they share nearly identical spectra and states from one system may be transformed into states of the other system by the action of differential operators known as supercharges. In one-dimensional quantum mechanics, supersymmetry has been used to study a plethora of new potentials including partners of: the Rosen-Morse potentials \cite{Compean_2005, Dominguez-Hernandez2011}, the truncated oscillator \cite{CMay2014, FERNANDEZC2018122, FernandezC.2019}, and the singular oscillator \cite{Marquette:2012sm}. In higher dimensional quantum systems, supersymmetry is less explored, though coherent states for the two-dimensional infinite well and its coordinate separable supersymmetric partners have been discussed \cite{Fiset_2015}. There is considerably more freedom when defining the supercharges for higher dimensional systems \cite{doi:10.1142/S0217732397000601}, and features which are exclusive to systems of dimension larger than one, such as non-coordinate separability of the Hamiltonian, can appear. In the present work we deal with precisely such a case where the initial Hamiltonian is coordinate separable, while its partner Hamiltonian is not \cite{PhysRevA.76.052114}.

The Morse potential was originally introduced to model anharmonic interactions in diatomic molecules which allow the possibility of the bond between the molecules breaking at sufficiently high energy \cite{PhysRev.34.57}. The uncoupled two-dimensional Morse potential has been used in the expansion of triatomic molecular interactions \cite{doi:10.1063/1.1670777, doi:10.1063/1.1670678, doi:10.1063/1.2140500, https://doi.org/10.1002/qua.21189}. More recently the Morse potential has been studied in the context of graphene \cite{Rode_2018, ZALI2021413045}, spectroscopy \cite{doi:10.1063/5.0013677}, and supersymmetry \cite{PhysRevA.76.052114, doi:10.1142/S0217751X1250073X}.

Describing coherent states for generalised quantum systems has been of interest since the canonical coherent states of the harmonic oscillator were formalised by Glauber and Sudarshan in their seminal works \cite{PhysRev.131.2766, PhysRevLett.10.277}. For the canonical coherent states there exist three equivalent definitions: as eigenstates of the annihilation operator; the orbit of the displacement operator acting on the vacuum; a particular superposition of Fock basis states \cite{Gazeau:2009zz}. For systems beyond the harmonic oscillator the three definitions typically do not coincide and one must choose the most applicable definition. Coherent states for two-dimensional harmonic oscillator systems with linear spectra have been discussed in \cite{isoand, 1259279089, Moran_2021} but in the present case we are dealing with a quadratically degenerate spectrum and must forego some of the nicer algebraic structures.

Generalised coherent states for the one-dimensional and two-dimensional Morse potentials have been studied in \cite{Angelova_2008} and \cite{Moran2021} respectively. For multidimensional quantum systems there exists a greater class of states when one departs from taking the tensor product of one-dimensional coherent states: the Hamiltonian may not be coordinate separable and thus its eigenstates may be entangled between its respective modes; the Hamiltonian may be coordinate separable but when defining a non-degenerate spectrum in order to define generalised coherent states \cite{PhysRevA.64.042104} the new basis states become entangled.

The paper is structured as follows: In section \ref{supersym} we review the supersymmetry formalism in two dimensions using second-order supercharges, after which we follow the work of Ioffe \cite{PhysRevA.76.052114} and compare the two partner Hamiltonians in section \ref{hamsec}. In section \ref{statesec} we develop set of non-degenerate states for the initial Hamiltonian which depend on two mixing parameters $\gamma_1, \gamma_2$, and transform allowed antisymmetric combinations of these states into a reduced set of non-degenerate states for the partner Hamiltonian. Following this in section \ref{sec5m} we construct an explicit example of these energy eigenstates for a value of the principle parameter $p=3\pi$, and then we develop coherent states for the partner Hamiltonian and discuss their properties in section \ref{csexamp}. Lastly we conclude with some open questions and remarks about the work.

\section{Second-order two-dimensional supersymmetry}\label{supersym}

Work on developing the supersymmetry formalism in two dimensions has been done by Ioffe and others \cite{IOFFE20062552, Ioffe_2004, Ioffe_2005}. In this context, the second order supercharges are given by
\begin{equation}\label{superop}
Q^+=(Q^-)^\dagger = g_{kl}(x,y)\hbar^2 \partial_k\partial_l +c_k(x,y)\hbar \partial_k +b(x,y), \quad k,l\in\left\{x,y\right\},
\end{equation}
where repeated indices are summed over and $g_{kl}(x,y), c_k(x,y)$ and $b(x,y)$ are real valued functions. Two partner Hamiltonians, $H, \tilde{H}$, are intertwined via
\begin{equation}\label{intert}
\tilde{H}Q^+=Q^+H, \quad HQ^-=Q^- \tilde{H}.
\end{equation}
Supposing we have the set of states of the initial Hamiltonian
\begin{equation}
\psi_{nm}(x,y)\equiv \bra{x,y}\ket{n,m},
\end{equation}
the implication of the intertwining relations \eqref{intert} is that for a given eigenstate $\psi_{nm}(x,y)$ of $H$, $Q^+ \psi_{nm}(x,y)$ is an eigenstate of $\tilde{H}$ with the same eigenvalue through
\begin{equation}\label{superproof}
H\ket{n,m}=E_{n,m}\ket{n,m} \implies Q^+H\ket{n,m}=E_{n,m}\left[Q^+\ket{n,m}\right]=\tilde{H}\left[Q^+\ket{n,m}\right],
\end{equation}
where we have suppressed the dependence on $x,y$ for brevity. In general (and in our example) there is not a one-to-one correspondence between states belonging to both Hamiltonians, certain eigenstates are not permissible because they become non-normalisable in the partner Hamiltonian \cite{IOFFE20062552, Ioffe_2004, Ioffe_2005}. That is to say while the intertwining relations \eqref{intert} always imply mathematically the eigenvalue equations \eqref{superproof} we may encounter a situation where the transformed eigenfunction, $Q^+\ket{n,m}$, is not physical.

In some cases additional normalisable eigenfunctions may be obtained for the partner Hamiltonian by transforming non-normalisable solutions of the initial Hamiltonian by the supercharge $Q^+$   \cite{DAS1999357, PANIGRAHI1993251}. In the present case, all normalisable solutions of the partner Hamiltonian were obtained by Ioffe \cite{PhysRevA.76.052114}.

General schemes for finding solutions for second order supercharges defined in \eqref{superop} do not exist. For the particular case of a metric, $g_{kl}(x,y)$, with Lorentz signature $(1,-1)$ the intertwining relations can be simplified and the coefficient appearing in \eqref{superop} can be found through a system of differential equations \cite{IOFFE20062552}.

Introducing the generators
\begin{equation}
\mathcal{Q}^+=\begin{pmatrix}
0 & Q^+ \\
0 & 0 
\end{pmatrix}, 
\quad \mathcal{Q}^-=\begin{pmatrix}
0 & 0 \\
Q^- & 0 
\end{pmatrix},
\quad \mathbf{H}=\begin{pmatrix}
\tilde{H} & 0 \\
0 & H 
\end{pmatrix},
\end{equation}\label{superalg}
we obtain the superalgebra defined by
\begin{equation}
[\mathbf{H},\mathcal{Q}^\pm]=0, \quad \{\mathcal{Q}^+,\mathcal{Q}^-\}=\mathcal{R},
\end{equation}
where $[\cdot,\cdot]$ and $\{\cdot,\cdot\}$ are the commutator and anticommutator respectively. The existence of the fourth order operator $\mathcal{R}$ is a consequence of defining second order supercharges \cite{ISI:000288078700002}. Comparing this with the implementation of supersymmetry using first order supercharges where the anticommutator of the supercharges is just the superHamiltonian.

\section{Initial and partner Morse Hamiltonians}\label{hamsec}
The separable two-dimensional Morse Hamiltonian in coordinate representation is given by
\begin{equation}\label{ham1}
H=-\frac{\hbar^2}{2m}\left(\partial_x^2 +\partial_y^2\right)+V_0\left(e^{-2\beta x}+e^{-2\beta y} -2 \left(e^{-\beta x}+e^{-\beta y} \right) \right)=H_x+H_y.
\end{equation}
Hamiltonian \eqref{ham1} permits a finite number of bound states which are solutions to the time independent Schr\"odinger equation $H\ket{n,m}=E_{n,m}\ket{n,m}$,
\begin{equation}\label{mors}
\psi_{n,m}\left(x,y\right)=\bra{x,y}\ket{n,m}=\mathcal{N}_{n,m}e^{-\frac{\tilde{x}}{2}-\frac{\tilde{y}}{2}}\tilde{x}^{p-n}\tilde{y}^{p-m}L_n^{2(p-n)}\left( \tilde{x}\right)L_m^{2(p-m)}\left( \tilde{y}\right).
\end{equation}
Here $L_n^\alpha(z)$ are the generalised Laguerre polynomials, the normalisation factor $\mathcal{N}_{n,m}$ is given by
\begin{equation}\label{n1fact}
\mathcal{N}_{n,m}=\beta \sqrt{\frac{(\nu-2n-1)(\nu-2m-1)\Gamma(n+1)\Gamma(m+1)}{\Gamma(\nu-n)\Gamma(\nu-m)}},
\end{equation}
and the tilde variables are defined as
\begin{equation}
 \tilde{x}=\nu e^{-\beta x}, \quad  \tilde{y}=\nu e^{-\beta y}.
\end{equation}
The parameter $\nu$ and an additional parameter $p$, are written in terms of the initial parameters by
\begin{equation}\label{mainparam}
\nu=\sqrt{\frac{8mV_0}{\beta^2 \hbar^2}}, \quad p=\frac{\nu-1}{2}.
\end{equation}

The bound state spectrum is given by
\begin{equation}\label{energy}
E_{n,m}=-\frac{\hbar^2 \beta^2}{2m}\left(\left( p-n\right)^2 +\left( p-m\right)^2\right).
\end{equation}

The spectrum is defined in such a way that states with negative energy are bound and states with zero or positive energy are unbound. For a discussion on the unbound states and spectrum, see, for example \cite{doi:10.1063/1.2364502, Gao-Feng2010, zhang2010morse}. The bound state conditions imply that the quantum numbers $n,m$ may take the following values \cite{alma991025163309705576}
\begin{equation}
n,m\in \left\{0,1,\ldots,\floor{p} \right\}.
\end{equation}
To facilitate counting arguments later, it is convenient to rewrite the parameter $p$ in terms of its integer part plus a remainder,
\begin{equation}\label{irrremainder}
p=k+\epsilon,\quad k=\floor{p},\quad \epsilon \in [0,1),
\end{equation}
where we further restrict to the case where $\epsilon$ is irrational to ensure that states are at most doubly degenerate \cite{Moran2021}.
 
The partner Hamiltonian we are interested in studying is given by \cite{PhysRevA.76.052114}
\begin{equation}\label{hampart}
\tilde{H}=H+\frac{\hbar^2 \beta^2}{2m \sinh^2 \left(\frac{\beta}{2}(x-y) \right)}.
\end{equation}
The two Hamiltonians \eqref{ham1} and \eqref{hampart} are related by \eqref{intert} through the supercharges
\begin{equation}\label{qplus}
Q^\pm=-H_x +H_y +D^\pm
\end{equation}
where $D^\pm$ is given by
\begin{equation}
D^\pm=\frac{\hbar^2 \beta^2}{2m}\coth \left(\frac{x-y}{2} \right) \mp \frac{\hbar^2\beta}{2m}\left( \left(\partial_{x}-\partial_y\right) + \coth \left( \frac{x-y}{2} \right)\left(\partial_{x}+\partial_y\right) \right).
\end{equation}
The supercharges $Q^\pm$ are related to a particular choice of \eqref{superop} where the metric $g_{kl}(x,y)$ has a Lorentz signature on the indices, $(1,-1)$, thus the second order derivative terms are not mixed between the two modes and there is a relative minus sign between $\partial_x^2$ and $\partial_y^2$.

So far we have kept factors of $\hbar$ to show exactly how the terms appear in the Hamiltonians and supercharges, but from here on we set the dimensionful quantities $\hbar=\beta=m=1$. The most striking features of the partner Hamiltonian \eqref{hampart} is its non-separability, and the non separable term is singular for $y=x$. 

\section{States of the initial and partner Morse Hamiltonians}\label{statesec}
Because the initial Hamiltonian is amenable to separation of variables its eigenstates are products of wavefunctions in each direction. In general, multidimensional quantum systems are degenerate in their energy spectrum, and in the view of addressing the construction of generalised coherent states it is important to have a non-degenerate increasing spectrum \cite{Klauder_1996, PhysRevA.64.042104}. For the finite quadratic spectrum \eqref{energy} the degeneracies are obtained via number theoretic means and, moreover, they are found to be at most doubly degenerate for irrational values of the parameter $p$. In \cite{Moran2021} general such degenerate combinations are constructed 
\begin{align}\label{nondeg}
\ket{\mu_i}_{p}&=
\begin{cases}
\gamma_1\ket{n,m}+\gamma_2\ket{m,n}, &\textrm{for }  \abs{n-m}>0,\\
\ket{n,n},  &\textrm{otherwise},
\end{cases}
\end{align}
 for complex coefficients $\gamma_1, \gamma_2$, satisfying the normalisability condition $\abs{\gamma_1}^2+\abs{\gamma_2}^2=1$. For the states $\ket{\mu_i}_{p}$, the index $i$ organises them in increasing energy order, and the subscript $p$ indicates the value of the parameter $p$ introduced in \eqref{mainparam}. 

Due to the singular nature of the partner Hamiltonian along the line $y=x$ we cannot construct states in the partner Hamiltonian with arbitrary coefficients. A detailed analysis provided in \cite{PhysRevA.76.052114} proves that the only way to compensate for this singularity to obtain normalisable states in the partner Hamiltonian is to transform perfectly antisymmetric combinations of the initial eigenfunctions in which case we obtain symmetric (about the line $y=x$) eigenfunctions in the partner Hamiltonian. In the notation we are using this corresponds to the choice of parameters $\gamma_1=-\gamma_2=\frac{1}{\sqrt{2}}$.

%\begin{align}\label{nondegspec}
%\ket{\mu_i}_{p}&=
%\begin{cases}
%\frac{1}{\sqrt{2}}\left(\ket{n,m}-\ket{m,n}\right), &\textrm{for }  \abs{n-m}>0,\\
%\ket{n,n},  &\textrm{otherwise},
%\end{cases}
%\end{align}
Additionally, Ioffe's analysis shows that the states admissible by the partner Hamiltonian require that $\abs{n-m}>1$ \cite{PhysRevA.76.052114}. This follows from the superalgebra \eqref{superalg} and specifically the existence of the fourth order operator $R=Q^-Q^+$. The operator $R$ commutes with the initial Hamiltonian, $[R,H]=0$ and we have that $R\ket{\mu_i}_p=r_{n,m}\ket{\mu_i}_p$ where
\begin{equation}\label{reval}
r_{n,m}=\frac{1}{2}\left((m-n)^2-1\right)\left((2p-m-n)^2-1\right),
\end{equation}
meanwhile the states in the partner system obtained by transforming the states of the initial system have norm given by \eqref{superproof}
\begin{equation}
\bra{\mu_i}_p Q^- Q^+ \ket{\mu_i}_p.
\end{equation}
We see immediately that the case where $n=m+1$ and vice versa lead to states in the partner Hamiltonian which are not normalisable because \eqref{reval} vanishes. The case where $n=m$ is excluded by definition when taking perfectly antisymmetric combinations of the eigenstates.

We take the convention that the positive coefficient appears in front of the term with $n>m$, thus we can exactly solve the partner Hamiltonian \eqref{hampart} to obtain the set of normalised states
\begin{align}\label{partner}
\ket{\nu_j}_p=\frac{1}{\sqrt{\mathcal{N}_j}}Q^+\ket{\mu_i}_{p}&=
\begin{cases}
\frac{Q^+}{\sqrt{2\mathcal{N}_j}}\left(\ket{n,m}-\ket{m,n}\right), &\textrm{for }  \abs{n-m}>1\\
\varnothing,  &\textrm{otherwise},
\end{cases}
\end{align}
where $\mathcal{N}_j=\abs{Q^+\ket{\mu_i}_p}^2$ are normalisation constants, $Q^+$ is defined in \eqref{qplus}, and $\varnothing$ means that nothing is to be done, i.e. do not generate an element for the set $\left\{\ket{\nu_j}_p\right\}$. The matching of the indices $i,j$ in \eqref{partner} is unique, but because the set $\left\{\ket{\mu_i}_{p}\right\}$ is larger than $\left\{\ket{\nu_j}_{p}\right\}$, the correspondence is not $i=j$. Due to the isospectrality given to us by the supersymmetry formalism \cite{COOPER1995267}, the states $\ket{\nu_j}_p$ are automatically arranged in increasing energy order because the initial states $\ket{\mu_j}_p$ were prepared in this way.

The Morse potential admits only a finite number of bound states, as such it is instructive to count the number of bound states in the initial and partner Hamiltonians. We know that the number of states in the set $\abs{\left\{ \ket{\mu_i}_p \right\}}$ is \cite{Moran2021}
\begin{equation}\label{set1}
\abs{\left\{ \ket{\mu_i}_p \right\}}=\frac{(k+1)(k+2)}{2},
\end{equation}
recalling that $k=\floor{p}$. From \eqref{partner} it is clear that the size of the set $\abs{\left\{ \ket{\nu_i}_p \right\}}$ must be smaller than \eqref{set1}. Imposing the condition $\abs{n-m}>1$, we count all pairs $(n,m)$.
\begin{table}[h]
\begin{center}
\begin{tabular}{ c|c } 
 $m$ & $n$ \\ 
 \hline
 $0$ & $2, \ldots, k$ \\ 
 $1$ & $3, \ldots, k$  \\ 
 $\vdots$ & $\vdots$  \\ 
 $k-2$ & $k$  \\ 
\end{tabular}
\caption{Allowed combinations of quantum numbers $(n,m)$ leading to normalisable states in the partner Hamiltonian.}
\label{tab1}
\end{center}
\end{table}
Adding up the possibilities across each row of table \ref{tab1}, we arrive at the following formula
\begin{equation}\label{set2}
\abs{\left\{ \ket{\nu_i}_p \right\}}=\sum_{j=1}^{k-1}(k-j)=\frac{k}{2}(k-1),
\end{equation}
thus the number of `missing' states between the two sets scales linearly in $k$:
\begin{equation}\label{miss}
\abs{\left\{ \ket{\mu_i}_p \right\}}-\abs{\left\{ \ket{\nu_i}_p \right\}}=1+2k.
\end{equation}
A key observation to be made is that bound states only exist in the partner Hamiltonian for $k\geq 2$ or equivalently, $p>2$.

Considering a scaled version of the spectrum \eqref{energy}
\begin{equation}\label{epsspec}
\varepsilon_{n,m}(p)=\varepsilon_{n,m}(k,\epsilon)=-\left[(p-n)^2+(p-m)^2\right], \quad n,m \in\left\{0,1,\ldots,\floor{p}\right\}.
\end{equation}
We recall that in \eqref{irrremainder} we write $p$ as the sum of its integer part plus an irrational remainder term, $\epsilon$, in which case the states of the initial Hamiltonian have a spectrum found by arranging the solutions of \eqref{epsspec} in increasing order
\begin{equation}
\varepsilon_{0,0}(k,\epsilon)<\varepsilon_{1,0}(k,\epsilon)=\varepsilon_{0,1}(k,\epsilon)<\ldots<\varepsilon_{k,k}(k,\epsilon),
\end{equation}
such that the spectrum for the states \eqref{nondeg} can be written as
\begin{equation}
\varepsilon_{\mu_0}(k,\epsilon)<\varepsilon_{\mu_1}(k,\epsilon)<\ldots<\varepsilon_{\mu_\textrm{max}}(k,\epsilon).
\end{equation}
Then by the isospectrality afforded to us by supersymmetry, we have the spectrum for the states \eqref{partner}
\begin{equation}\label{nuspecarr}
\varepsilon_{\nu_0}(k,\epsilon)<\varepsilon_{\nu_1}(k,\epsilon)<\ldots<\varepsilon_{\nu_{\textrm{max}}}(k,\epsilon).
\end{equation}
We remark that $\mu_\textrm{max}$ and $\nu_{\textrm{max}}$ are determined by \eqref{set1} and \eqref{set2} respectively.
\section{Application to $p=3\pi$}\label{sec5m}
We will now apply the formalism set up in the preceding sections to the example of $p=3\pi$. It is worth mentioning that while the analysis of the degeneracies relies on the irrationality of the principle parameter, $p$, in order to implement calculations on a computer algebra system we do need to approximate $p$ by a rational number. That being said, as long as states are at most double degenerate (which is easily verified on a computer), the analysis still holds.
\subsection{Energy eigenstates}\label{stateexamp}
We begin by recounting the non-degenerate energy eigenstates for the initial system for $p=3\pi$. The example used in \cite{Moran2021} gives  us the following $55$ state set, $\mathcal{S}$, for the initial Hamiltonian
\begin{equation}\label{setofs}
\mathcal{S}=\left\{ \ket{\mu_0}_{3\pi}, \ket{\mu_1}_{3\pi}, \ldots, \ket{\mu_{54}}_{3\pi}\right\}, \quad \abs{\mathcal{S}}=55,
\end{equation}
where
\begin{align}\label{states}
\ket{\mu_i}_{3\pi}&=
\begin{cases}
\ket{0,0}, & i=0\\
\gamma_1\ket{1,0}+\gamma_2\ket{0,1}, & i=1\\
\gamma_1\ket{2,0}+\gamma_2\ket{0,2}, & i=2\\
\ket{1,1}, & i=3\\
\vdots \\
\ket{9,9}, & i=54.
\end{cases}
\end{align}
Following this, we set $\gamma_2=-\gamma_1=-\frac{1}{\sqrt{2}}$ and determine the $36$ state set for the partner Hamiltonian, $\tilde{
\mathcal{S}}$, using \eqref{partner},

\begin{equation}\label{setofstilde}
\tilde{\mathcal{S}}=\left\{ \ket{\nu_0}_{3\pi}, \ket{\nu_1}_{3\pi}, \ldots, \ket{\nu_{35}}_{3\pi}\right\}, \quad \abs{\tilde{\mathcal{S}}}=36.
\end{equation}
States in the set $\tilde{\mathcal{S}}$ preserve orthonormality by construction. Explicitly the first few and last states are
\begin{align}\label{states}
\ket{\nu_i}_{3\pi}&=
\begin{cases}
\frac{1}{\sqrt{\mathcal{N}_0}}Q^+\left[\frac{1}{\sqrt{2}}\left(\ket{2,0}-\ket{0,2}\right)\right], & i=0\\
\frac{1}{\sqrt{\mathcal{N}_1}}Q^+\left[\frac{1}{\sqrt{2}}\left(\ket{3,0}-\ket{0,3}\right)\right], & i=1\\
\frac{1}{\sqrt{\mathcal{N}_2}}Q^+\left[\frac{1}{\sqrt{2}}\left(\ket{4,0}-\ket{0,4}\right)\right], & i=2\\
\frac{1}{\sqrt{\mathcal{N}_3}}Q^+\left[\frac{1}{\sqrt{2}}\left(\ket{3,1}-\ket{1,3}\right)\right], & i=3\\
\vdots \\
\frac{1}{\sqrt{\mathcal{N}_{35}}}Q^+\left[\frac{1}{\sqrt{2}}\left(\ket{9,7}-\ket{7,9}\right)\right], & i=35,
\end{cases}
\end{align}
where the normalisation constants $\mathcal{N}_i=\abs{Q^+\ket{\mu_k}}^2$ where the index $i$ in the set $\tilde{\mathcal{S}}$ is to be identified with the state with index $k$ in the set $\mathcal{S}$ from which it is generated. In addition we verify that the difference in size of the two sets of states satisfy \eqref{miss} for $k=9$.

As with the initial separable two-dimensional Hamiltonian, the organisation of the spectrum and states of the partner Hamiltonian is algorithmic, so there is no obvious pattern to be seen. This follows from the ordering of numbers as the sum of two squares. And since the partner Hamiltonian is isospectral to the initial Hamiltonian, the partner Hamiltonian also inherits this problem with the additional complication related to the restriction $\abs{n-m}>1$ discussed in the previous section. Nevertheless, organising the states in this way can be done quite easily with most computer algebra systems.

Consider a scaled version of the spectrum \eqref{nuspecarr} where the constant term $-2\epsilon^2$ has been removed,
\begin{equation}\label{specex}
\tilde{\varepsilon}_{\nu_i}(9, 0.42478)=-\left[(9-n)^2 +(9-m)^2+2(0.42478)\left(18-n-m\right)\right],
\end{equation}
arranged in increasing order such that $\abs{n-m}>1$.
\begin{figure}[H]
\centering
\includegraphics[scale=1.2]{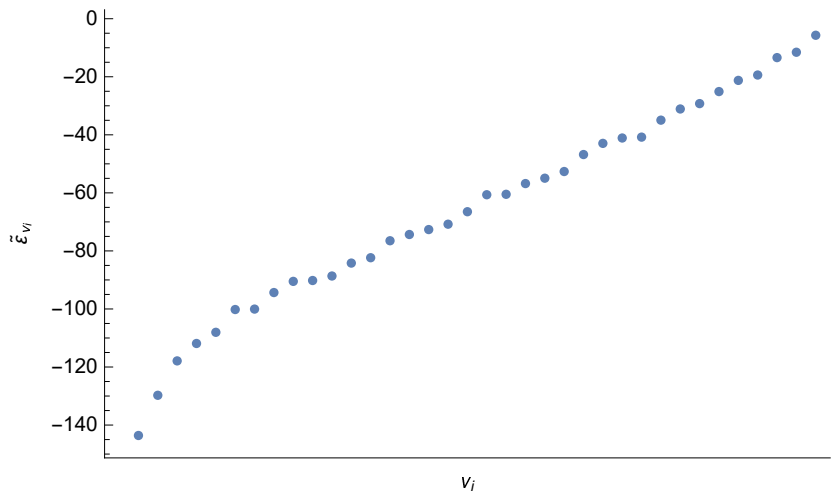}
\caption{Energies $\tilde{\varepsilon}_{\nu_0},\ldots,\tilde{\varepsilon}_{\nu_{35}}$ for $p=3\pi$ arranged in increasing order.}
\label{egraph}
\end{figure}
In figure \ref{egraph} we see the spectrum of the partner Hamiltonian arranged in increasing order. Notice that there does not appear to be a nice functional form to the spectrum, this is because the techniques required to understand the degeneracy of an arbitrary energy level are number theoretical and in general must be computed on a case by case basis.
\begin{figure}[H]
    \centering
    \subfloat{{\includegraphics[scale=0.65]{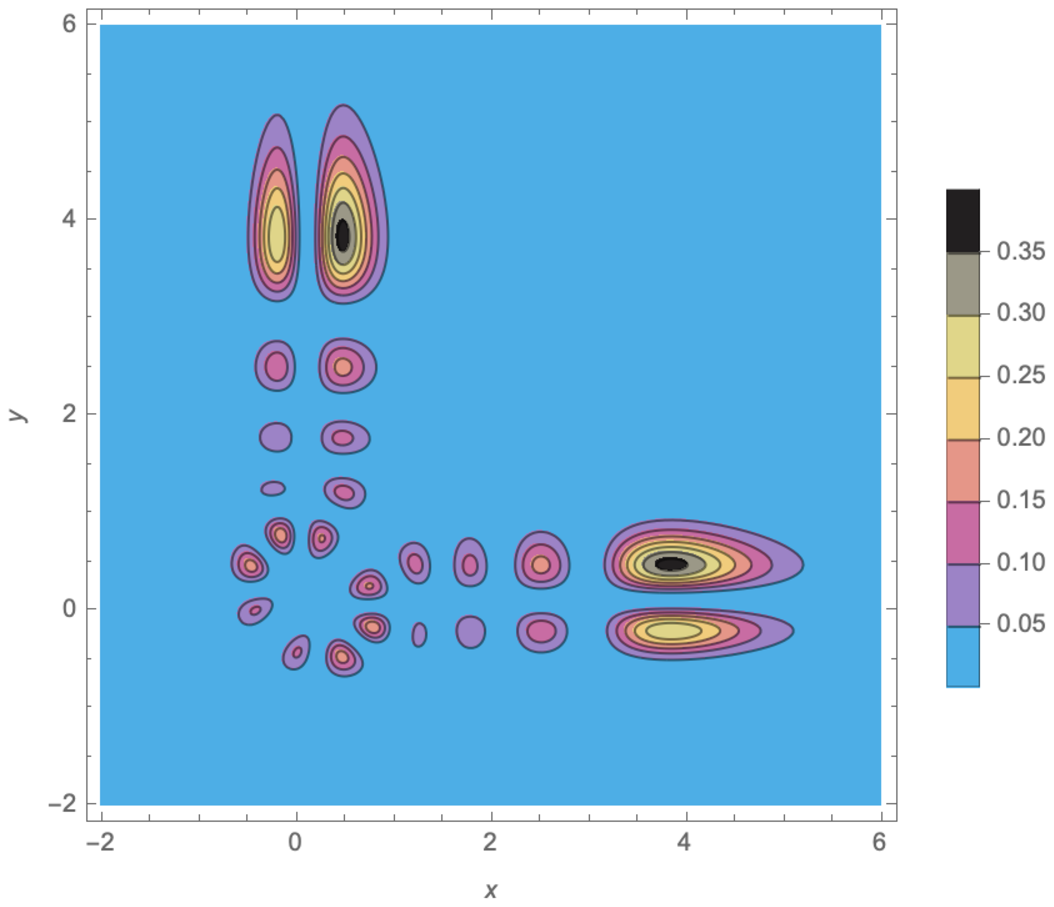} }}%
    \qquad
    \subfloat{{\includegraphics[scale=0.65]{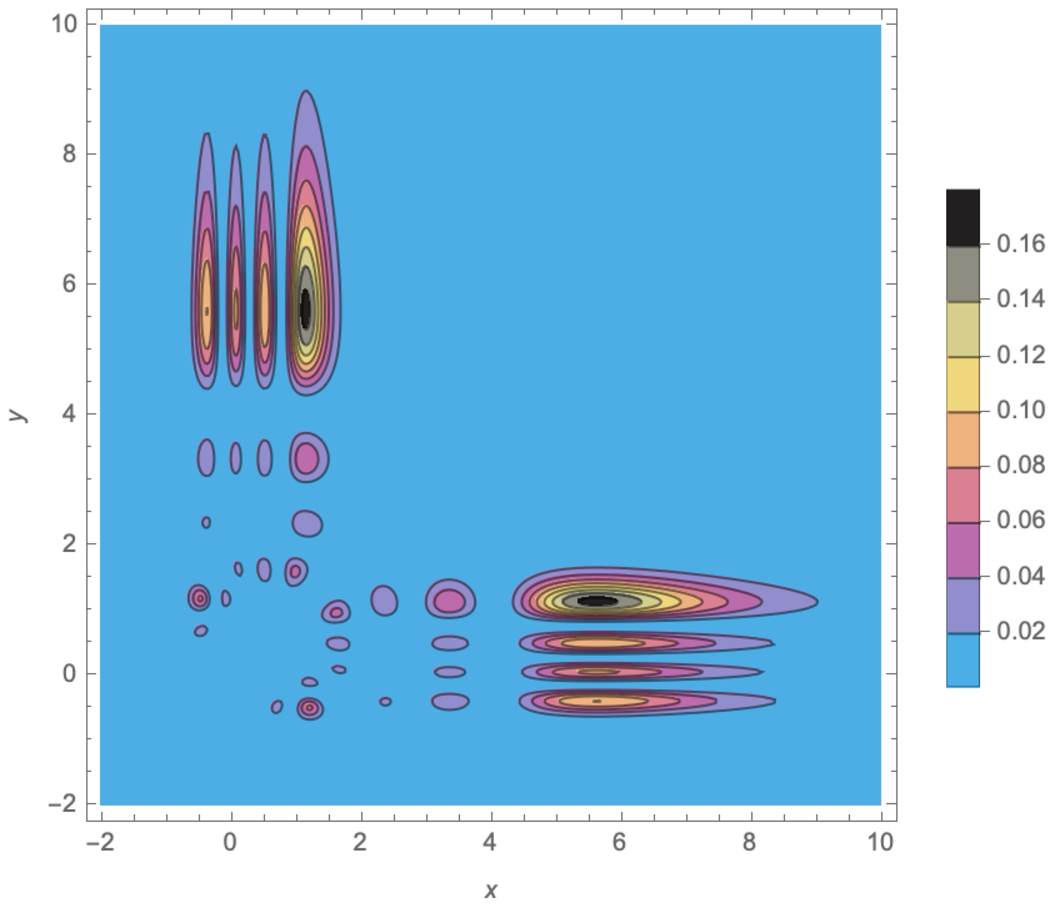}}}%
    \caption{Probability densities of $ \abs{\bra{x,y}\ket{\nu_{15}}_{3\pi}}^2$ (left) and $ \abs{\bra{x,y}\ket{\nu_{25}}_{3\pi}}^2$ (right).}%
    \label{fig22}%
\end{figure}

We see in figure \ref{fig22} that the configuration space wavefunctions $\bra{x,y}\ket{\nu_i}_{3\pi}$ (and their probability densities) are symmetric and non-singular about the line $y=x$. The eigenstates $\ket{\nu_i}$ have similar patterns in their wavefunctions to those of the original Hamiltonian $\ket{\mu_j}$ \cite{Moran2021}, the key differences being that the wavefunction $\bra{x,y}\ket{\nu_i}_{3\pi}$ is always identically zero on the line $y=x$ and they are always symmetric about the line too. In contrast, the parameters $\gamma_1,\gamma_2$ in $\bra{x,y}\ket{\mu_i}_{3\pi}$ are free up to a normalisation and allow us to adjust the symmetry or antisymmetry of the wavefunctions. Additionally we find that for larger values of $i$, $\abs{\bra{x,y}\ket{\nu_i}_{3\pi}}^2$ contain more detailed structure in their probability densities and have more `islands' of non-zero probability density, while for smaller $i$ there is less detail to the structure and fewer `islands' of probability.
\subsection{Coherent states}\label{csexamp}
Let us suppose the existence of some ladder operator on the set $\tilde{\mathcal{S}}$,
\begin{equation}
\begin{split}
&\mathcal{B}^+\ket{\nu_i}_p = \sqrt{f(i+1)} \ket{\nu_{i+1}}_p,\\
&\mathcal{B}^-\ket{\nu_i}_p = \sqrt{f(i)} \ket{\nu_{i-1}}_p,
\end{split}
\end{equation}
where the functions $f(i)$ must be positive and satisfy suitable boundary conditions. In practice the $f(i)$ are often taken to be the relative difference of the energy to the ground state energy.

We can construct generalised coherent states as approximate eigenstates of $\mathcal{B}^-$,
\begin{equation}
 \mathcal{B}^-\ket{\Phi}_p \approx \Phi \ket{\Phi}_p.
\end{equation}
Expanding the generalised coherent states $\ket{\Phi}_{p}$ for $p=3\pi$ in the basis \eqref{setofstilde} we find the following
\begin{equation}\label{gencos}
\ket{\Phi}_{3\pi}=\frac{1}{\sqrt{\mathcal{N} (\Phi)}}\sum_{n=0}^{35}\frac{\Phi^n}{\sqrt{[\tilde{\varepsilon}_{\nu_n}-\tilde{\varepsilon}_{\nu_0}]!}}\ket{\nu_n}.
\end{equation}
We remark that the state $\ket{\Phi}_{p}$ is only an approximate eigenstate of $\mathcal{B}^-$ because the term proportional to $\ket{\nu_\textrm{max}}$ does not appear in the expansion \eqref{gencos}, we add it in by hand. This is a general feature of ladder operator eigenstate definitions of coherent states with finite spectra.
%\begin{figure}[H]
   % \centering
    %\subfloat{{\includegraphics[scale=0.66]{z001.eps} }}%
    %\qquad
    %\subfloat{{\includegraphics[scale=0.66]{z01.eps}}}%
    %\caption{$ \abs{\bra{x,y}\ket{\Phi}_{3\pi}}^2$, with $\Phi=0.001$ (left) and $\Phi=0.01$ (right)}%
    %\label{fig:1}%
%\end{figure}
\begin{figure}[H]
    \centering
    \subfloat{{\includegraphics[scale=0.66]{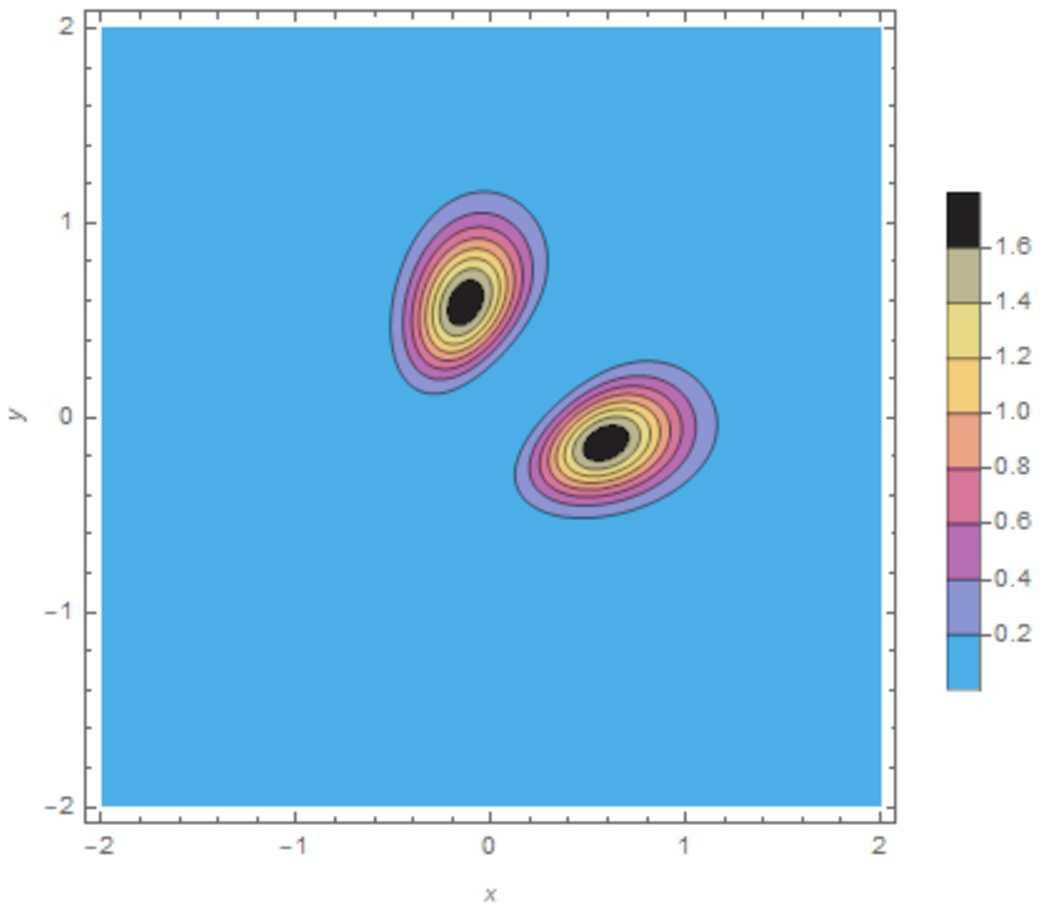} }}%
    \qquad
    \subfloat{{\includegraphics[scale=0.66]{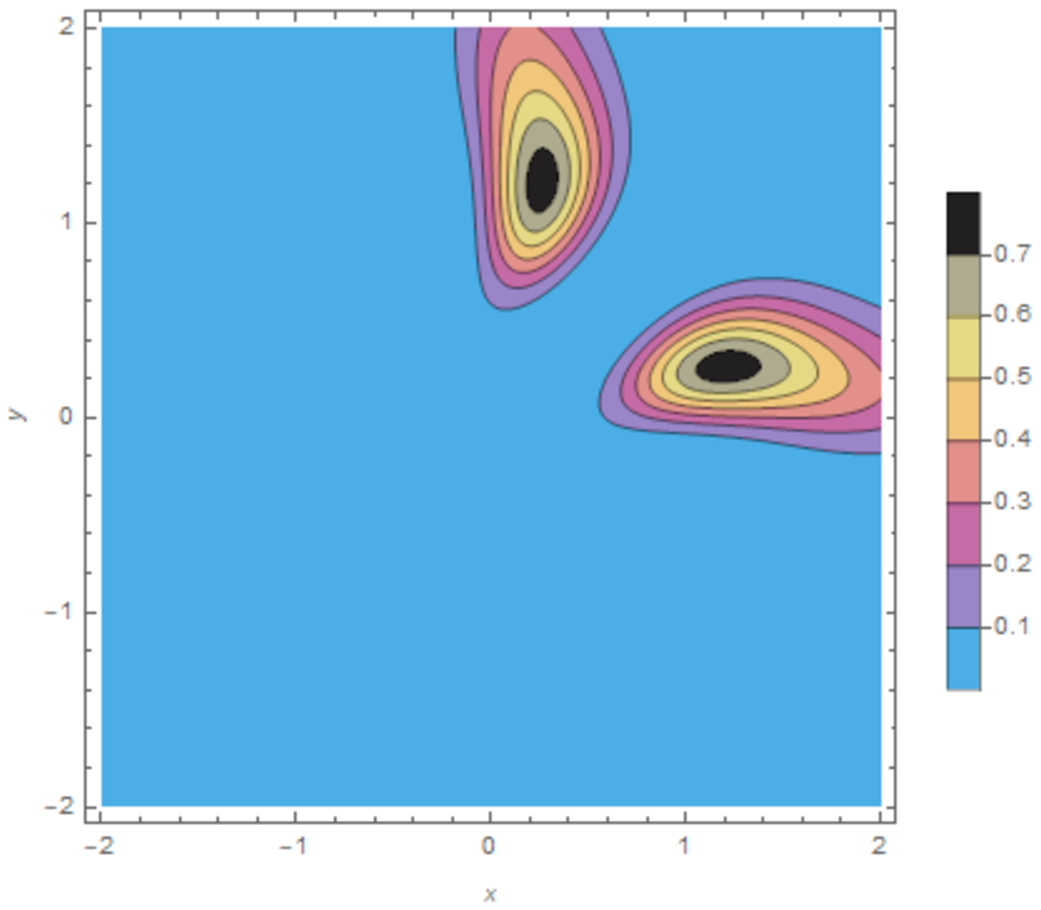}}}%
    \caption{$ \abs{\bra{x,y}\ket{\Phi}_{3\pi}}^2$, with $\Phi=0.001$ (left) and $\Phi=5$ (right)}%
    \label{fig:2}%
\end{figure}
The behaviour of the generalised coherent states is qualitatively very different to the original 2D Morse problem \cite{Moran2021}. A result of the non separability of the potential means that the probability density concentrates on two symmetric peaks around the line $y=x$. Even if we let $\Phi$ become smaller, this separation remains.

Moreover we can also analyse the the uncertainty relations for the coherent states. The variance of an operator, $\hat{O}$, is defined by
\begin{equation}
 \left( \Delta \hat{O} \right)^2=\langle \hat{O}^2 - \langle \hat{O}\rangle^2 \rangle,
\end{equation}
and since we are working in the position representation, the momentum and position operators (in the $x$ direction), as well as the Heisenberg uncertainty relation take the form
\begin{equation}
 \hat{P}_x =-i \frac{\mathrm{d}}{\mathrm{d}x}, \quad \hat{Q}_x =x, \quad \left(\Delta \hat{Q}\right)^2\left(\Delta \hat{P}\right)^2\geq \frac{1}{4}.
\end{equation}

For canonical coherent states of the harmonic oscillator the variance in the position and momentum quadratures is equal, $\left(\Delta \hat{Q}\right)^2=\left(\Delta \hat{P}\right)^2=\frac{1}{2}$. Whenever $\left(\Delta \hat{Q}\right)^2\not=\left(\Delta \hat{P}\right)^2$ we have squeezing between the quadrature operators, furthermore when either quadrature has variance smaller than $\frac{1}{2}$ we have sub-shot-noise squeezing.
\begin{figure}[H]
    \centering
\includegraphics[scale=1.2]{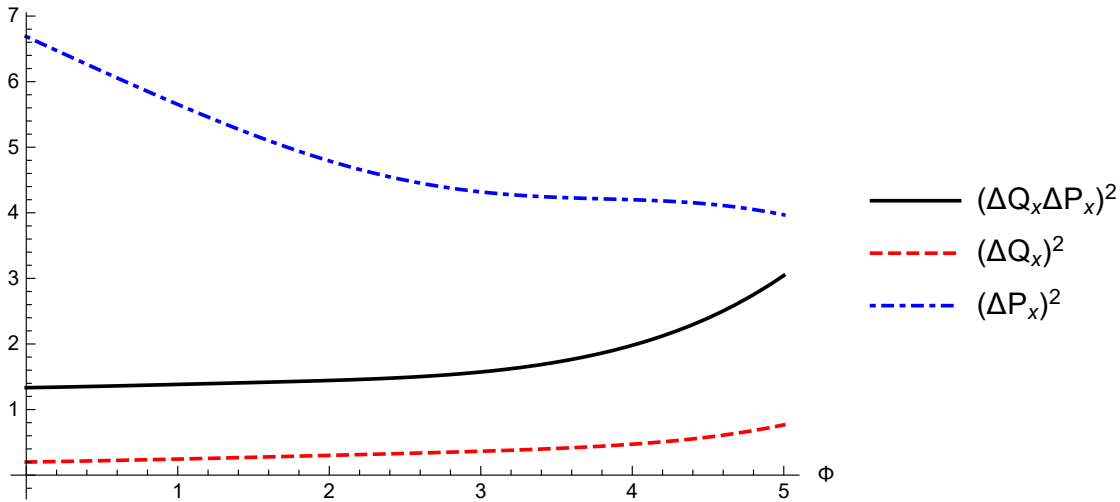} 
    \caption{Uncertainty in the position and momentum quadratures and their product.}%
    \label{fig:2uncert}%
\end{figure}
No matter the value of $\Phi$, the generalised coherent states $\ket{\Phi}_p$ never minimise the Heisenberg uncertainty relation. This is due to the existence of the singularity on the line $y=x$. In quantum physics the notion of non-classicality is used to explain purely quantum phenomena (see, for example, \cite{PhysRevLett.122.040503, Zelaya2017, PhysRevA.102.032413}). In the present case we have a few markers of non-classicality: we have eigenstates of a non-separable Hamiltonian which are themselves non-separable, they cannot be written as the tensor product of two one-dimensional states $\ket{\Phi}_p\not=\ket{\phi_x}\otimes\ket{\phi_y}$; the states are never minimal uncertainty; there is significant squeezing between the position and momentum quadratures. 

In figure \ref{fig:2uncert} we see that the squeezing in the position quadrature is below that of the canonical coherent states, that is $\left(\Delta \hat{Q}_x\right)^2<\frac{1}{2}$. In the language of quantum optics this refers to sub-shot-noise squeezing, these techniques allow one to obtain higher resolution imaging by reducing the uncertainty in quadrature at the expense of increasing it in the conjugate quadrature \cite{PhysRevD.23.1693, Abadie2011, PhysRevLett.78.3105, Li2021}. This indicates that despite the fact that the coherent states $\ket{\Phi}_p$ are unable to localise onto a single point in space, there is little statistical variance in the accuracy of its position. Other examples of states which appear to localise at two spatially separated points include the well studied non-classical cat states \cite{Duan2019}.
%\section{Ladder operators \textcolor{red}{Not worthwhile}}
%Previously we suppose that there exists a pair of ladder operators $\mathcal{B}^+, \mathcal{B}^-$ on the initial set $\mathcal{S}$ such that
%\begin{equation}
%\begin{split}
%&\mathcal{B}^+\ket{\mu_i}_p = \sqrt{f(i+1)} \ket{\mu_{i+1}}_p\\
%&\mathcal{B}^-\ket{\mu_i}_p = \sqrt{f(i)} \ket{\mu_{i-1}}_p
%\end{split}
%\end{equation}
%The function $f(i)$ must satisfy the boundary conditions $f(0)=f(\xi+1)=0$. We know that we cannot transform every state into a normalisable state in the new set $\tilde{\mathcal{S}}$, but we can transform a restricted subset. We may build a generalised set of ladder operators on the non-separable Hamiltonian. We know that
%\begin{equation}
%\begin{split}
%&Q^+\ket{\mu_i} \propto \ket{\nu_j},\\
%&Q^+\ket{\mu_k} \propto \ket{\nu_l},
%\end{split}
%\end{equation}
%if $i<k$ this also implies $j<l$. Possibly due to the 'unpredictable' distribution of the energies and quantum numbers it is not straightforward to implement a ladder operator on the new set of states $\tilde{\mathcal{S}}$ with some combination of $Q^+, \mathcal{B}^\pm$ as
%\begin{equation}
%\mathcal{Q}^+ \mathcal{B}^+
%\end{equation}
\section{Conclusion}
In this paper, following the work of Ioffe \cite{PhysRevA.76.052114}, we constructed an explicit set of eigenstates for the non-separable singular two-dimensional Morse potential. We were able to connect these states with a set of non-degenerate states for the initial separable Morse potential \cite{Moran2021}. We constructed coherent states for the non-separable Morse Hamiltonian and found that the configuration space wavefunction is unable to localise at the origin due to the singularity present in the potential. This extends work previously done in the domain of multidimensional coherent states. Not only are the states entangled, but they arise from a non-coordinate separable Hamiltonian which is singular along the line $y=x$ resulting in strongly non-classical behaviour. The procedure we developed in this paper is algorithmic and in principle can be used to describe coherent states for any two-dimensional system with quadratically degenerate spectrum and its supersymmetric partners (if they exist).

Continuing the analysis we computed the uncertainty relation for the coherent states and while the states themselves were not minimal uncertainty, significant squeezing was found between the position and momentum quadratures with sub-shot-noise squeezing in the position quadrature. This indicates strong quantum behaviour similar to that of the highly non-classical cat states. The statistical variance in the position quadrature is smaller than that of the canonical coherent states yet the wavefunction appears to localise onto two space-like separated regions. For systems of dimension larger than two, one should expect richer structure still, and the emergence of multipartite entangled systems.

Lastly, we remark that a more detailed study of the coherent states for interacting multidimensional quantum systems, such as the triatomic molecular Hamiltonian \cite{doi:10.1063/1.1670777} and the Pais-Uhlenbeck oscillator \cite{doi:10.1142/S0219887816300154}, would be of interest going forward. It is clear that we can expect novel behaviour that might not be found from non-interacting multidimensional generalisations, though in practice their solution is much more difficult to obtain.
\section*{Acknowledgements}
J. Moran acknowledges the support of the D\'epartement de physique at the Universit\'e de Montr\'eal. V. Hussin acknowledges the support of research grants from NSERC of Canada. Both authors would like to thank I. Marquette for his help in the preparation of this paper.
%%%%%%%%%%%%%%%%%%%%%%%%%%%%%%%%%%%%%%%%%%%%%%%%%%%%%%%%%%%%%%%%%%%%%%%%%%%%%%%%%%%%%%%
%%Bibliography%%
\bibliographystyle{aip}
\bibliography{refs.bib}

\end{document}